\documentstyle[12pt]{article}
\def\beqa{\begin{eqnarray}}
\def\eeqa{\end{eqnarray}}
\def\beq{\begin{equation}}
\def\eeq{\end{equation}}

\def\dmunu{_{\mu\nu}}

\def\bib#1{$^{\ref{#1}}$}

\let\alp=\alpha

\def\pl{{\it Phys. Lett.}\ }

\def\apj{{\it Ap. J.}\ }
\def\aj{{\it Astron. J.}\ }

\def\ie{{\it i.e. }}

\def\p{\phi}

\def\dis{\displaystyle}

\begin{document}
\def\bib#1{[{\ref{#1}}]}

\begin{titlepage}
	 \title{Luminosity variation  in (de)focusing microlensing}

\author{{S. Capozziello$^{1,2}$, R.de Ritis$^{1,2}$, V. I. Mank'o$^{3,4}$,
A.A. Marino$^{1,2}$, G. Marmo$^{1,2}$}
\\ {\em {\small $^{1}$Dipartimento di Scienze Fisiche, Universit\`{a} di 
Napoli,}}
\\ {\em {\small $^{2}$Istituto Nazionale di Fisica Nucleare, Sezione di 
Napoli,}}\\
   {\em {\small Mostra d'Oltremare pad. 19 I-80125 Napoli, Italy,}}\\
 {\em {\small $^{3}$Osservatorio Astronomico di Capodimonte,}}\\
{\em {\small Via Moiariello 16 I-80131 Napoli, Italy}}\\
{\em {\small $^{4}$Lebedev Physical Institute, Leninsky Pr., 53, Moscow 117924, 
Russia.}}}

	      \date{}
	      \maketitle

  \begin{abstract}
The luminosity variation of a stellar source due to
the gravitational microlensing effect can be considered also if the light
rays are defocused (instead of focused) toward the observer.
In this case, we  should detect a gap instead of a peak in the light curve
of the source. Actually, we describe how the phenomenon depends on the 
relative position
of source and lens with respect to the observer: if the lens is between,
we have focusing, if the lens is behind, we have defocusing.
It is shown that the number of events with predicted gaps is equal to the
number of events with peaks in the light curves.

 \end{abstract}	      

\vspace{20. mm}
e--mail address:\\
capozziello@axpna1.na.infn.it\\
deritis@axpna1.na.infn.it\\
manko@astrna.na.astro.it\\
marino@axpna1.na.infn.it\\
gmarmo@axpna1.na.infn.it

	      \vfill
	      \end{titlepage}

\section{\normalsize \bf Introduction}
Microlensing is a specific application of gravitational lensing which 
is mainly used to search for the  
{\it Massive Astrophysical Compact Halo Objects} (MACHOs) \bib{paczynski}, 
which are today considered as the most probable 
constituents of the dark halo of
 our Galaxy \bib{carr}, \bib{padmanabhan}; however other possibilities are today
explored. In \bib{gurevich1},\bib{gurevich2},\bib{gurevich3}, microlensing
by cold dark matter particles and noncompact objects is considered.

The term
 "microlensing" is used since
 the angular separation of  the two
images, usually produced  by a point--like mass lens, is  too small 
to be resolved
($\sim 10^{-6}$ arcsec). 
However, even if it is not possible to detect  multiple images, the
magnification can still be seen when the lens and the source move 
relatively to each other: this motion gives rise to 
a lensing--induced time variability
of the source luminosity \bib{chang}. 
The effect was firstly observed for  quasars 
\bib{irwin},\bib{schild}, so that we have to distinguish {\it galactic}
 and {\it extragalactic}  microlensing.
In the first case, the light sources are stars and the angular separations
involved are $\sim 10^{-3}$arcsec, in the second case, the sources are
 quasars and the angular separations involved are
$\sim 10^{-6}$arcsec. The term "microlensing" is used in both cases. 
The principle on which the phenomenon  is based is the following. 
If the closest approach between a point mass 
lens and a source is equal or less  than the Einstein angular radius
$\theta_{E}$, 
the peak magnification in lensing--induced light curve 
corresponds to a brightness
enhancement which can be easily  detected with the today facilities
\bib{paczynski}. 
The Einstein angular radius $\theta_{E}$
is a feature of the system lens--source--observer 
which furnishes the natural angular 
scale to describe the lensing geometry. Starting from the gravitational
lens equation \bib{ehlers}, it is given by
\beq
\label{0.3}
\theta_{E}=\sqrt{\frac{4GM(\leq r_{E})D_{ls}}{c^{2}D_{ol}D_{os}}}\,,
\eeq
where $D_{ls}, D_{ol}, D_{os}$ are respectively the distances
lens--source, lens--observer, and source--observer.
The angular distance $\theta_{E}$
corresponds to the effective distance
$ r_{E}=\theta_{E}D_{ol}$. 
 The symbol $M(\leq r_{E})$ means that the mass of the lens
has to be contained inside a sphere whose radius is the Einstein one.
 For multiple imaging, 
$\theta_{E}$ gives the typical angular separation among the single images; 
for axisymmetric lens--source--observer systems, it gives the aperture of a
circular bright image, called {\it Einstein ring}.
 Sources which are 
closer than $\theta_E$ to the optical axis experience strong lensing effect 
and are hardly magnified, sources which are located well outside of 
the Einstein ring are not very much magnified. In other words, 
for a lot of lens models, 
the Einstein ring represents the boundary between the zones where sources
are strongly magnified or multiply--imaged and those where they are softly
magnified or singly--magnified  \bib{ehlers}.

In order to detect microlensing, (and then MACHOs)
the first proposal \bib{paczynski} was to monitor millions of stars 
in the
{\it Large Magellanic Cloud}
(LMC), or in the bulge of Galaxy in order to look for such magnifications. 
If enough events are detected, it should
be possible to map the distribution of (dark) mass objects in 
the halo  of 
Galaxy or between the Solar System and the bulge of Galaxy.
Due to the distances involved, both approaches can be used 
for "galactic microlensing"
\bib{paczynski},\bib{alcock}.
The expected time scale for microlensing--induced luminosity 
variations is given in 
terms of the typical angular scale $\theta_{E}$, the relative velocity $v$
between source and lens, and the distance of the observer to the lens 
$D_{ol}$, that is $\Delta t=(D_{ol}\theta_{E})/v$.
If light curves are sampled with time intervals between  the hour and the
year, the mass range of MACHOs is $10^{-6}\div 10^{2} M_{\odot}$,
that is from planets to very massive stars (or black holes).
These numbers are in agreement with theoretical constraints 
\bib{padmanabhan},\bib{ehlers}.
The chance of seeing microlensing events depends on the 
{\it optical depth}, which is the probability that at any instant  a given 
source is within the angle $\theta_{E}$ of a lens, 
as we shall see below.

Several groups are searching for MACHOs 
\bib{alcock},\bib{aubourg}, \bib{udalski},\bib{alard}
but so far,  
 few experimental data  (about 100 events) 
 can be considered
statistically relevant in order to allow to draw conclusions on the
physical properties of MACHOs like their mass.
 
An important point has to be discussed.
Until now, microlensing is  considered for lenses which  focus light rays
toward the observer.
On the other hand, in optics, there exists the opposite
effect if the refraction index of media is appropriately chosen
and if the relative positions of the source and the lens is changed 
with respect to the observer. That is, we can
ask the question whether there exist or not distributions of matter 
producing gravitational fields which deflect the light rays in a manner which
mimics defocusing lenses of usual optics. 
The wish to introduce and to study the notion of 
{\it defocusing gravitational lens} is mainly motivated by 
the hypothesis that the 
microlensing events with luminosity peak may be accompained by the
existence of events with valley in the luminosity curve. This inverse 
phenomenon can be easily  understood by the following considerations.
In the "standard" studied situation,  a
MACHO is between the source and the observer and the   emitted rays  
 are slightly curved in the direction of the observer and such a
fact produces the a luminosity magnification.  The
opposite situation is statistically as probable as the previous one when a
MACHO is located behind the source with respect to the observer.  Then, the 
source rays are slightly curved out of the observer direction 
which  detects a gap in  luminosity. 
We will discuss precisely this situation using the equations for 
geodesics (along which light rays move) 
in a generic Schwarzschild gravitational field.

In this paper, we discuss  the (de)focusing microlensing considering
a simple model in which a MACHO moves with respect to the source and the
observer. However, the discussion can be generalized in  a statitistical
way by considering several sources and lenses.

\section{\normalsize \bf Luminosity variation  induced by a  point mass lens}

If $\theta_{s}$ is the angular size of the source and the condition 
$\theta_{s}\leq\theta_{E}$  holds, the magnification due to the
microlensing effect must be $\mu\geq 1.34$, 
(see, for example, \bib{ehlers},\bib{refsdal}).
 A magnification $\mu\sim 1.34$ corresponds to
a magnitude enhancement of $\Delta m\sim 0.32$.
In other words, we can say that when the true position of a light
source lies inside the Einstein ring, the total magnification of the two
images that it yields amounts to $\mu\geq 1.34$. This means that the angular 
cross
section for having significant microlensing effects (\ie $\mu\sim 1.34$ and 
$\Delta m \sim 0.32$),  is equal to $\pi\theta_{E}^{2}$.
Such a cross section,  from
(\ref{0.3}), is proportional to the mass $M$ of the deflector and to
the ratio of the distances involved.
Such considerations allow to calculate the {\it optical depth}.

Let us consider the case of randomly distributed point--mass lenses:
it is possible to estimate the frequency of significant gravitational lensing 
events from the observations of distant compact sources, that is we are
considering optical systems where the involved angular sizes  are 
much smaller than $\theta_{E}$. In this situation, the magnification of a 
compact source is equal or greater than $1.34$ 
(since $\theta_{s}<\theta_{E}$) and the probability $P$ to have a  
significant microlensing event for a randomly located 
compact source at a distance $D_{os}$ is given by
\beq
\label{0.14}
P=\frac{\pi\theta_{E}^{2}}{4\pi}=\left(\frac{D_{ls}}{D_{os}D_{ol}}\right)
\left(\frac{GM}{c^{2}}\right)\,.
\eeq
Such a probability is linear in 
the mass $M$ of deflector so that it holds also when several 
point--mass lenses are
acting. Assuming a constant density 
for the lens(es) and a static background (this last assumption
surely holds for galactic distances), averaging opportunely  on the distances
$D_{ls}, D_{ol}, D_{os}$, the probability (\ref{0.14}) can be interpreted as 
the  optical depth $\tau$   for lensing 
\bib{refsdal},\bib{harwit},\bib{press}: 
\beq
\label{optical}
P=\tau=-\left(\frac{D_{ls}}{D_{os}}\right)\frac{U}{c^{2}}\,,
\eeq
where
$ U=-GM/D_{ol}$,
is the Newton potential due to the lens as measured by the observer.
In other words, $\tau$ corresponds to the fraction of sky covered by the 
Einstein ring. 
Due to the fact that the deflecting masses change the
path of light rays,  the observer will detect different luminosities
for a given source when the deflector is present and when it is not
present: then,  the optical depth  is related with such a
relative luminosity change as we shall see below.

Before discussing how to realize (de)focusing, we have to consider the motion
of light ray paths in a gravitational field in order to obtain the luminosity
variations due to the presence of point mass lenses.
We have to take into account the 
geometric optic approximation since we are assuming 
that  light propagates as rays.

As it is  known
\bib{ehlers}, 
 a gravitational field has the same effect of a
medium  in which light rays propagates. 
For weak gravitational fields, 
the metric tensor components $g\dmunu$ can be expressed in terms of
 Newton gravitational potential $\Phi$. 
The 
refraction index $n$, in this case,   is related to the 
gravitational potential $\Phi({r})$ produced by some matter distribution,
that is
$n=1-2\Phi(r)/c^{2}\,$.
If the rays pass near a spherical body of mass $M$, they will 
undergo the action
of a Schwarzschild gravitational field described by the metric element
\beq
\label{2.3}
ds^{2}=\left(1-\frac{R_s}{r}\right)c^{2}t^{2}-
\frac{dr^{2}}{{\left(1-\frac{R_s}{r}\right)}}-
r^{2}\left(d\theta^{2}+\sin^{2}\theta d\phi^{2}\right)\,,
\eeq
where
$R_{s}=2MG/c^{2}$
is the Schwarzschild radius.
The light ray trajectories passing near the deflecting body can be easily
found by solving the problem of motion connected with (\ref{2.3}).
If the condition $r\gg R_{s}$ holds (that is the light ray passes well
outside of the surface where the metric becomes singular), the trajectory is
\beq
\label{2.10}
r=r_{0}\left\{\cos(\phi-\phi_{0})+
\frac{R_s}{2r_{0}}\left[2-\cos^{2}(\phi-\phi_{0})\right]\right\}^{-1}\,,
\eeq
which is nothing else but a straight line
corrected by a
hyperbolic--like term in polar coordinates 
\bib{papapetrou}. 
The parameters
$r_{0}$ and $\phi_{0}$ are the initial data of the problem; $r_{0}$
 is the distance of the line from
the origin of coordinates, $\p_{0}$ is a given angle which tells us how much
the line is tilted with respect to the polar axis.  
The deflecting mass is 
set at the origin of reference frame. The amount of  deviation
from the rectilinear behaviour depends on the ratio 
$R_s/r_{0}$, that is on the mass $M$ of the 
gravitational source and on the  parameter $r_{0}$.
In Cartesian coordinates
Eq.(\ref{2.10}) becomes
\beq
\label{2.17}
r_{0}=Ax+By+\left(\frac{R_s}{r_{0}}\right)\sqrt{x^{2}+y^{2}}-
\left(\frac{R_{s}}{2r_{0}}\right)\frac{(Ax+By)^{2}}{\sqrt{x^{2}+y^{2}}}\;,
\eeq
where 
$A=\cos\phi_{0}$ and $B=\sin\phi_{0}$.

Let us consider now the limit
$r\rightarrow\infty$. 
Eq.(\ref{2.10}) becomes an algebraic equation for $\cos(\p-\p_{0})$ from
which
 we get (being $r_{0}\gg R_{s}$)
\beq
\label{2.11}
\cos(\phi-\p_{0})\simeq -\frac{2MG}{c^{2}r_{0}}\,
\eeq
which indicates how the  gravitational field $(M\neq 0)$ deviates
the rays from the straight line direction. If
$M=0$ or $r_{0}\rightarrow\infty$ (that is in absence of gravitational field
or when  $r_{0}$ is very large), we have
$\cos(\phi-\p_{0})=0\,,$ that is $\phi-\p_{0}=\pm\frac{\pi}{2}\,.$
If the gravitational field is weak, in the limit 
$r\rightarrow\infty$, we have
$\phi-\p_{0}=\pm(\delta+\pi/2)$,
from which, by substituting into (\ref{2.10}), we get
$\sin\delta\simeq \delta=R_s/r_{0}$
being $\delta$ small. The total amount of ray deviation gives the standard
result
$2\delta\simeq 4MG/(c^{2}r_{0}),$
which is  the deflection angle due to a point mass 
acting as a gravitational lens.

Now, taking in mind such a results, 
 we want to obtain the general formula describing the 
 variation of   luminosity of a radiation source
in the sky induced by  a gravitational microlensing effect.
We will show that such a variation is due to the change of
 direction of  light rays (geodesics)  which move in a given  
nonstationary matter distribution.
In other words, we are
supposing that a given
background metric $g^{(1)}\dmunu$ is modified by a passing heavy body
(a MACHO) which locally perturbs it so that we have to consider  a new metric
$g^{(2)}\dmunu$. The effect of such a background change is a deviation in the
direction of geodesics which  gives 
 a bundle of hyperbolic--like curves (instead of the unperturbed
 bundle of straight lines).
Two cases are possible:
 the observable variation
of source luminosity is due  to a microlensing focusing effect
or to  a microlensing
defocusing effect. 
In the first case, a MACHO  is between the source
and the observer producing focusing, that is, at a certain moment,
the alignment $(I)$
$$
\mbox{source}\,\;\;\longrightarrow\;\;\,\mbox{lens}\,\;\;
\longrightarrow\;\;\,\mbox{observer}\,,
$$
is realized; in the second case, a MACHO 
 is behind the source and light rays are defocused
toward the observer, that is, at a certain moment the alignment $(II)$
$$
\mbox{lens}\,\;\;\longrightarrow\;\;\,\mbox{source}\,\;\;\longrightarrow
\;\;\,\mbox{observer}\,,
$$
is realized.
In the first case, the observer detects an increasing
 luminosity, in the second case, he detects a decreasing one.
The problem can be easily formulated by a geometric model
in which, given a reference frame, 
we assign the position of the light source and the position of 
the detector (a telescope)
in a  background metric $g^{(1)}\dmunu$. Then we calculate
the geodesics which give the light--ray paths. Furtherly, considering a 
MACHO passing
between the source and the observer or behind the source (with respect to the
observer), the metric becomes locally $g^{(2)}\dmunu$ and 
the geodesics will change
giving focusing or defocusing of light rays. 
Let us  start by choosing a system of Cartesian
coordinates. We put the source in
$(x_{S},\, y_{S})=\{-a,\,0\}$  in the case of configuration $(I)$
and $(x_{S},\, y_{S})=\{ a,\,0\}$  in the case of configuration 
$(II)$.
The telescope is   in
$(x_{T},\, y_{T})=(R,\,h)\,$ and we are assuming that the MACHO passes in
the origin $(0,\,0)$. At the beginning, it is not present and the metric is 
$g^{(1)}\dmunu$.
There exists a unique light ray (a unique geodesic) 
which intersects the source and the
upper limit of the  telescope aperture.

Let us now suppose that, due to a redistribution of matter,
the metric becomes $g^{(2)}\dmunu$,
(the simplest case is to consider  a passing MACHO). 
This event modifies the structure of geodesic bundle  from the source to 
the observer. Schematically, we have  a different bundle 
geodesics between the source
and the upper limit of the aperture of the telescope. 
The rays which reach the upper limit of  the telescope 
in the  metric $g^{(1)}\dmunu$ are emitted at the angle $\alpha_{1}$, while 
they are  emitted at the angle $\alpha_{2}$ in the metric
$g^{(2)}\dmunu$.
In the first case,  each geodesic is given by a function
$y_{1}(x)$ in Cartesian coordinates; in the second one by 
a function $y_{2}(x)$. The angles
$\alp_{1}$ and $\alp_{2}$ are given by the derivatives
\beq
\label{pippo}
\tan\alp_{1}=\frac{dy_{1}}{dx}\left|_{(\mp a;0)}\right.\,,
\;\;\;\;\;\;\;\;\;\;\; 
\tan\alp_{2}=\frac{dy_{2}}{dx}\left|_{(\mp a;0)}\right.\,,
\eeq
calculated in the coordinates of the source.
The variation of luminosity is related to the variation of
the direction of geodesics, that is to the change of the number of light rays 
which reach the telescope, so that
the general formula for the relative
change of luminosities in both cases is
\beq
\label{3.11}
\frac{\Delta L}{L}=
\pm\left[\left(\frac{dy_{2}(x)}{dx}-
\frac{dy_{1}(x)}{dx}\right)
\left(\frac{dy_{1}(x)}{dx}\right)^{-1}\right]_{({x_{S}},{y_{S}})}^{2}\,,
\eeq
where
the two  derivatives of geodesics are calculated in the coordinates of the 
source.

Let us now apply these general considerations
to the case of a flat metric which is
perturbed by the gravitational field of a moving MACHO.
This means that the initial metric $g^{(1)}\dmunu$ is a 
 Minkowski one while the perturbed metric $g^{(2)}\dmunu$ is
the Schwarzschild one.
Without MACHO, geodesics are straight lines emitted by the source,
that is 
${\dis r=r_{0}[\cos(\p-\p_{0})]^{-1}}$,
in polar coordinates, or
$ r_{0}=Ax+By$,
in Cartesian coordinates. The constants $A$ and $B$ are the same 
as above. When a MACHO (passing in the origin) perturbs the background,
the geodesics are given by Eq.(\ref{2.10}) (or (\ref{2.17})). By calculating
the derivative in the position of the source (that is in $\{\pm a,0\}$)
 and using (\ref{3.11}), we get
\beq
\label{3.18'}
\frac{\Delta L}{L}
=\pm\left(\frac{R_s}{2r_{0}}\right)^{2}
\left\{\frac{A^{2}+2}{A\left[1\pm
A\left(\frac{R_s}{r_{0}}\right)\right]}\right\}^{2}\,,
\eeq
where plus sign means "focusing" and then a peak in light curve of the
source detected by the observer, while minus sign means "defocusing" and then
a gap in the light curve detected.
Eq.(\ref{3.18'}) (by Eqs.(\ref{pippo}))  
shows that the variation of luminosity 
depends on  the relative 
positions of the lens  and 
the light ray ($r_{0}$, $\p_{0}$),  on the mass of the MACHO $M$,
as well as on the relative position of the lens and the
light source ($\{\pm a,0\}$). 

Such calculations can be performed in any  configuration of the system
 source--lens--observer.
 Here, for simplicity, we have taken into account source, lens
and observer lying on the same line. 

\section{\normalsize \bf The mass of the lens and the optical depth}
Using the above formulas, we can  estimate the mass 
of of a  MACHO acting as a lens
both for focusing and defocusing cases. From Eq.(\ref{3.18'}),
 we have
\beq
\label{3.21}
M=\left(\frac{c^{2}r_{0}}{G}\right)
\left[\frac{A\sqrt{|\Delta L/L|}}{2+
A^{2}\mp 2A^{2}\sqrt{|\Delta L/L|}}\right]\;,
\eeq
where now minus sign refers to focusing and plus to defocusing.
The modulus tells us 
that both the peak
and the gap in light curve give indications on the MACHO mass.

By  Eqs.(\ref{0.14}) and (\ref{optical}), the optical depth is
\beq
\label{depth}
\tau_{\pm}= \left(\frac{D_{ls}}{D_{ol}}\right)
\left(\frac{r_{0}}{D_{ol}}\right)
\left[\frac{A\sqrt{|\Delta L/L|}}{2+
A^{2}\mp 2A^{2}\sqrt{|\Delta L/L|}}\right]\;,
\eeq
where $\tau_{+}$ is the optical depth (probability) connected to a focusing
event while $\tau_{-}$ is associated to a defocusing one.

In order to give some numbers, we  obtain a MACHO of mass 
$M\sim 0.5\div 1 M_{\odot}$, 
if $\Delta L/L\sim 10^{-2}$,
$r_{0}\simeq r_{E}$ of the order of one astronomic unit  and 
$ \p_{0}\sim |\delta|+\pi/2$,
with $ |\delta|\sim 10^{-5}$. Such result holds for
focusing and defocusing MACHOs. On the other hand, it is easy to 
obtain the optical
depth  $\tau\sim 10^{-6}$ 
 toward the Galactic bulge
and $\tau\sim 10^{-7}\div 10^{-8}$ 
toward the LMC \bib{paczynski},\bib{alcock}. 
The similar results
are also obtained for 
if $\Delta L/L\sim 10^{-4}$ and
$|\delta|\sim 10^{-3}$. 
In principle, we can cover all
the  mass range $10^{-6} M_{\odot}$ 
to $10^{2} M_{\odot}$    expected for MACHOs.  
However,
we have to stress that, statistically, the features of the light curves 
are not expected to be
completely symmetric: in fact, for a randomly chosen focusing and defocusing 
configurations $(I)$ and $(II)$ in which
the distances involved are the same, we can estimate that if the peak
magnification is, for example, of a factor 3, 
the gap for defocusing is of the order 0.1.

\section{\normalsize \bf Conclusions}
We have pointed out that   microlensing effects could be detected
not only if we observe peaks in luminosity curve of  sources, but
also if we detect  gaps. However, our prediction is that for observed
to date quantity of events with given peak magnifications, must exist 
approximately the same quantity of events with gaps in light curve.

Furthermore, by the  knowledge of  the geometry (and the relative positions)
of the optical  system source--lens--observer,
we can estimate both the mass and the optical depth for a given lens.

 These facts could contribute to bypass one of the lack of
microlensing detecting experiments: that is the low number of observed events
(till now about 100,  not all exactly tested, 
for millions of detected source stars).
Roughly speaking,
 one could
expect to double the number of succesful detections including
also defocusing events.

It is worthwhile to note that when several MACHOs are present, the 
previous discussion still holds due to the Fermat principle (see, for example
\bib{ehlers}). The effect is additive and it is similar to that of a 
light ray passing
through different media with refraction indexes $n_{1},\cdots n_{j}$.
Then, in principle, it is possible to evaluate the total deviation of
a light ray  by summing up the effects 
of the various deflectors.

Finally, we have to stress that in a statistical approach to the
microlensing, 
our approach gives rise to two contributions to the
number density $n(D_{l})$ of lenses, one coming from focusing objects
$n_{+}(D_{l})$ and another coming from defocusing ojects $n_{-}(D_{l})$. 

\vspace{4. mm}

{\bf ACKNOWLEDGMENTS}\\
The authors want to thank Michal Jaroszy\'{n}ski for the useful
discussions on the topic. V.I.M. thanks the Director of the 
Osservatorio Astronomico  di Capodimonte for the hospitality during 1996.

\begin{center}
{\bf REFERENCES}
\end{center}

\begin{enumerate}
\item\label{paczynski} 
B. Paczy\'{n}ski, \apj {\bf 301}, 503 (1986);
 \apj {\bf 304}, 1 (1986);
 {\it Gravitational Lenses} 
Lecture Notes in Physics {\bf 406}, p. 163, Springer--Verlag, Berlin (1992).
\item\label{carr}
B.J. Carr, in {\it Proceedings of Int. Workshop on the Identification
of Dark Matter}, Sheffield 8--12 Sept. 1996 (World Scientific).
\item\label{padmanabhan}
T. Padmanabhan, {\it Structure Formation in the Universe},
Cambridge Univ. Press, Cambridge (1993).
\item\label{gurevich1}
A.V. Gurevich, V.A. Sirota, and K.P. Zybin, \pl {\bf 207 A}, 333 (1995).
\item\label{gurevich2}
A.V. Gurevich K.P. Zybin, \pl {\bf 208 A}, 276 (1995).
\item\label{gurevich3}
A.V. Gurevich,  K.P. Zybin,  and V.A. Sirota, \pl {\bf 214 A}, 232 (1996).
\item\label{chang}
K. Chang and S. Refsdal, {\it Nature} {\bf 282}, 561 (1979).
\item\label{irwin}
M.J. Irwin et al. \aj {\bf 98}, 1989 (1989).
\item\label{schild}
R.E. Schild and R.C. Smith, \aj {\bf 101}, 813 (1991).  
\item\label{ehlers}
P.V. Blioh and A.A. Minakov {\it Gravitational Lenses}
Kiev, Naukova Dumka (1989) (in Russian);
P.Schneider, J. Ehlers, and E.E. Falco {\it Gravitational Lenses}
Springer--Verlag, Berlin (1992);
R. Kaiser {\it Gravitational Lenses} 
Lecture Notes in Physics {\bf 404}, p. 143, Springer--Verlag, Berlin (1992). 
\item\label{alcock}
C. Alcock et al., {\it Nature} {\bf 365}, 621 (1993);
C. Alcock, R.A. Allsman, T.S. Axelrod et al., \apj {\bf 461}, 84 (1996).
\item\label{aubourg}
E. Aubourg et al., {\it Nature} {\bf 365}, 623 (1993).
\item\label{udalski}
A. Udalski et al., {\it Acta Astron.} {\bf 42}, 253 (1992).
\item\label{alard}
C. Alard in: {\it Astrophysical Applications of Gravitational Lensing}
Proc. IAU Symp. (1995)
173, eds. C.S. Kochanek and J.N. Hewitt (Boston, Kluwer)
\item\label{refsdal}
S. Refsdal, {\it Proc. Intern. Conf. on Rel. Theories of Gravitation},
London (1970);
 \apj {\bf 159}, 357 (1970);
S. Refsdal and J. Surdej, {\it Rep. Prog. Phys.} {\bf 57}, 117 (1994).
\item\label{harwit}
M. Harwit, {\it Astrophysical Concepts}, Ed. Springer, Berlin (1988).
\item\label{press}
W.H. Press and J.E. Gunn, \apj {\bf 185},  397 (1973).
\item\label{papapetrou}
A. Papapetrou, {\it Lectures on General Relativity}, Reidel Pub. Company,
Boston (1974).
\end{enumerate}
\vfill

 \vfill

\end{document}